\documentclass[twocolumn,aps,prl,showpacs,amsmath,amssymb,superscriptaddress,floatfix]{revtex4-1}
\usepackage[english]{babel}
\usepackage{graphicx}
\usepackage{dcolumn}
\usepackage{bm}
\usepackage{hyperref}
\usepackage[mathlines]{lineno}
\usepackage{xcolor}
\usepackage{tikz}
\usetikzlibrary{patterns}
\usepackage{relsize}
\usepackage{comment}
\usepackage{csquotes}

\usepackage{xcite}
\usepackage{xr}
\makeatletter
\newcommand*{\addFileDependency}[1]{ 
  \typeout{(#1)}
  \@addtofilelist{#1}
  \IfFileExists{#1}{}{\typeout{No file #1.}}
}
\makeatother

\newcommand*{\myexternaldocument}[1]{%
    \externaldocument[SM-]{#1}%
    \addFileDependency{#1.tex}%
    \addFileDependency{#1.aux}%
}
\myexternaldocument{Supplementary_Material/supplementary_material}

\usepackage[normalem]{ulem}

\usepackage{natbib}
\usepackage{braket}

\begin{document}



\title{Lattice polarons across the superfluid to Mott insulator transition}
\author{V. E. Colussi}
\email[Corresponding author: ]{colussiv@gmail.com}
\affiliation{INO-CNR BEC Center and Dipartimento di Fisica, Universit\`a di Trento, Via Sommarive 14, 38123 Povo, Trento, Italy}
\author{F. Caleffi}
\affiliation{International School for Advanced Studies (SISSA), Via Bonomea 265, I-34136 Trieste, Italy}
\author{C. Menotti}
\affiliation{INO-CNR BEC Center and Dipartimento di Fisica, Universit\`a di Trento, Via Sommarive 14, 38123 Povo, Trento, Italy}
\author{A. Recati}
\affiliation{INO-CNR BEC Center and Dipartimento di Fisica, Universit\`a di Trento, Via Sommarive 14, 38123 Povo, Trento, Italy}
\affiliation{Trento Institute for Fundamental Physics and Applications, INFN, Via Sommarive 14, 38123 Povo, Trento, Italy}

\date{\today}

\begin{abstract}
We study the physics of a mobile impurity confined in a lattice, moving within a Bose-Hubbard bath at zero temperature. Within the Quantum Gutzwiller formalism, we develop a beyond-Fr\"ohlich model of the bath-impurity interaction. Results for the properties of the polaronic quasiparticle formed from the dressing of the impurity by quantum fluctuations of the bath are presented throughout the entire phase diagram, focusing on the quantum phase transition between the superfluid and Mott insulating phases. Here we find that the modification of the impurity properties is highly sensitive to the different universality classes of the transition, providing an unambiguous probe of correlations and collective modes in a quantum critical many-body environment.
\end{abstract}

\maketitle


\textit{Introduction.} --
Polarons, quasiparticles formed by a mobile impurity dressed by a cloud of virtual excitations of the bath in which it is immersed, are ubiquitous in physics. The most famous examples include lattice polarons in semiconductors, magnetic polarons in strongly-correlated systems \cite{franchini2021}, and $^3$He atoms in superfluid $^4$He \cite{bardeen1967}.
Recent pioneering realizations of polarons using highly imbalanced mixtures of ultracold gases have provided a new platform to study the role of the bath in determining the quasiparticle properties \cite{PhysRevLett.102.230402, PhysRevLett.103.170402, PhysRevLett.117.055301, PhysRevLett.117.055302}. A major advantage of cold gases is that the bath is clean and, importantly, its equation of state and coupling with the impurity can be controlled. In this context, impurities immersed in a non-interacting Fermi sea or a weakly-interacting Bose gas have been extensively explored both experimentally and theoretically (see e.g. \cite{scazzaReview22} and refs. therein).

\begin{figure}[!t]
\centering
\includegraphics[width=1\linewidth]{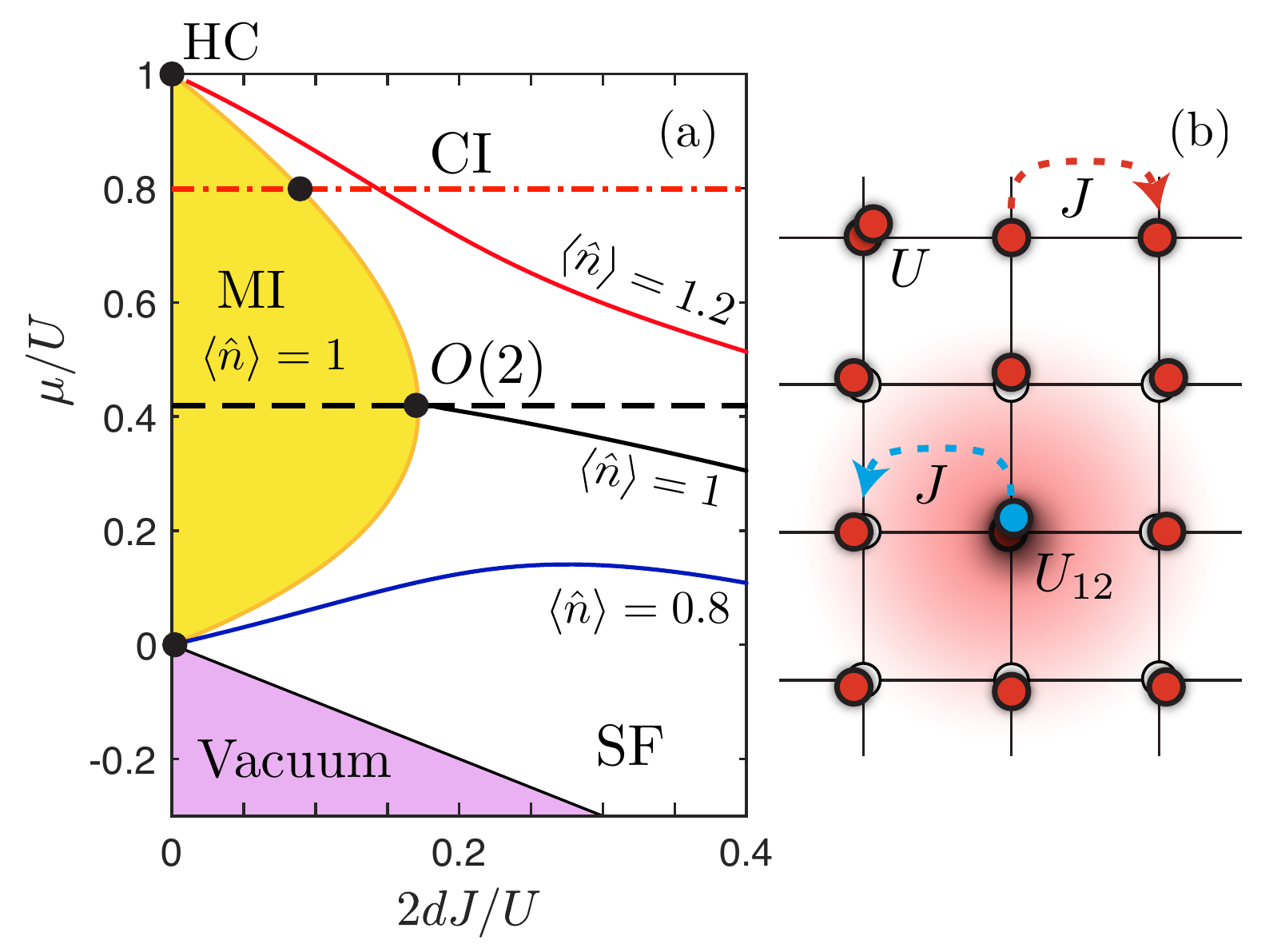}
\caption{(a) Gutzwiller phase diagram of a $d$-dimensional Bose-Hubbard bath around the $\langle \hat{n} \rangle = 1$ Mott lobe showing constant $\langle \hat{n} \rangle$ (solid) and $\mu/U$ (dashed and dashed-dotted) lines. The Mott insulator (MI) to superfluid (SF) transition can be crossed at the tip via the $O(2)$ transition or on the edges via the commensurate-incommensurate (CI) transition. For increasing interaction strength $U$, non-integer filling lines connect deep and hard-core (HC) regimes of the SF. (b) A mobile impurity of equal mass with hopping $J$ and coupling $U_{12}$ to a BH bath is dressed by a cloud of excitations, producing a lattice Bose polaron.}
\label{fig:1}
\end{figure}

From a complementary point of view, an impurity in a bath constitutes an open quantum system.  It has been realized that ultracold atomic platforms are also ideal to study such systems in non-trivial situations \cite{streif2016, Mitchison2016, cosco2018}. 
Of particular interest is the case when the bath undergoes a phase transition. The impurity acts as a probe for the phase transition itself, with its properties significantly modified by the presence of a critical point, as discussed previously for fixed (or infinite mass), magnetic-like impurities \cite{caleffi2021}. In particular, within the decoherence model for a Bose-Hubbard (BH) bath \cite{rossini2007}, it was shown in \cite{caleffi2021} that the impurity dephasing dynamics is strongly affected by the superfluid (SF) to Mott insulator (MI) phase transition of the bath \cite{greiner2002}, having a qualitatively different behavior depending on whether the transition is crossed at the commensurate-incommensurate (CI) or the $O(2)$ (fixed density) critical point [see Fig.~\ref{fig:1}(a)]. 
For a mobile impurity, the well-defined quasiparticle pole was observed to disappear close to the transition temperature for Bose-Einstein condensation \cite{yan2020science}. Recently, the prospect of using the energy of a mobile impurity as a probe for the finite temperature superfluid transition in a Fermi gas was explored theoretically in \cite{Alhyder22}. 

In this Letter, we consider a mobile impurity coupled to a BH bath in order to determine the properties of the lattice Bose polaron across the MI/SF phase transition. We study the problem using the quantum Gutzwiller (QGW) approach that we have recently developed \cite{caleffi2018, colussi2021quantum}. This method allows the bath-impurity interaction to be recast in terms of elementary excitations, in a form which can be seen as an extension of the well-known Fr\"olich model for polarons in crystals \cite{frohlich54}. This extension includes both the many-band structure of the spectrum and of the nonlinear couplings, which are fundamental to describe the transition to the MI phase. The polaron properties are then determined using second-order perturbation theory. In particular, we find that the impurity properties show a non-monotonic behavior close to the critical points and a strong sensitivity to the universality class of the transition.

\textit{Model.} --
We consider a mobile impurity coupled to a two-dimensional BH model, on a uniform square lattice composed by $I$ sites with lattice spacing $a$. The microscopic Hamiltonian is $\hat{H} = \hat{H}_B + \hat{H}_I + \hat{H}_{IB}$, where
\begin{align}\label{eq:model}
    &\hat{H}_B = -J \sum_{\langle {\bf r, r'} \rangle} \hat{a}^\dag_{\bf r} \, \hat{a}_{\bf r'} + \frac{U}{2} \sum_{\bf r} \hat{n}_{\bf r} \left( \hat{n}_{\bf r} - 1 \right) - \mu \sum_{\bf r} \hat{n}_{\bf r} \, , \\
    &\hat{H}_I = -J \sum_{\langle {\bf r, r'} \rangle} \hat{a}^\dag_{I, \bf r} \, \hat{a}_{I, \bf r'} \, , \qquad \hat{H}_{IB} = U_{12} \sum_{\bf r} \hat{n}_{\bf r} \, \hat{n}_{I, {\bf r}} \, , \nonumber
\end{align}
are the Hamiltonians describing the bath, the impurity and the bath-impurity coupling, respectively. The bosonic operators $\hat{a}^\dag_{\bf r}$ and $\hat{a}_{\bf r}$ create and destroy, respectively, a bath particle at lattice site ${\bf r}$ and are related to the local density operator as $\hat{n}_{\bf r} = \hat{a}^\dag_{\bf r} \, \hat{a}_{\bf r}$. Here, $\varepsilon_{\bf k} = 4 \, J \sum_{i = 1}^{d} \sin^2(k_i \, a/2)$ is the $d$-dimensional free-impurity dispersion, $\hat{a}^\dag_{I,{\bf r}}$ ($\hat{a}_{I, {\bf r}}$) and $\hat{n}_{I, {\bf r}} = \hat{a}^\dag_{I,{\bf r}} \, \hat{a}_{I,{\bf r}}$ are the creation (destruction) and local density operators of the impurity, while $U$ ($\mu$) is the on-site interaction strength (chemical potential) of the bath. For simplicity, we assume that the hopping energy between neighboring bath sites $J$ and the bare effective mass $M^{-1} = 2 \, J \, a^2$ of the impurity are equal, working in units $\hbar = 1$. Only weak impurity-bath on-site couplings $U_{12}$ are considered, such that the impurity is a probe with negligible back action on the zero-temperature ground state of the bath.

In order to fully characterize the BH bath across the phase diagram shown in Fig.~\ref{fig:1}(a), we employ the recently developed QGW method \cite{PhysRevResearch.2.033276}. This approach has the advantage of providing a robust, semi-analytical description of both local and non-local quantum correlations in BH models, showing remarkable agreement with Quantum Monte Carlo predictions even in critical regimes where quantum fluctuations are strong~\cite{PhysRevResearch.2.033276, SciPostPhys.12.3.111}. Fluctuations $\delta \hat{c}_n({\bf r})$ on top of the mean-field Gutzwiller ground state $\bigotimes_{\bf r} \sum_n c^0_n \, \left| n, {\bf r} \right\rangle$ \cite{PhysRevB.45.3137, PhysRevA.84.033602, KRUTITSKY2016} are quantized in terms of the elementary many-body excitations of the system $\delta \hat{c}_n({\bf r}) = I^{-1/2} \sum_{\alpha{\bf k}} e^{i \, {\bf k} \cdot {\bf r}} \, (u_{\alpha, {\bf k}, n} \, \hat{b}_{\alpha,{\bf k}} + v_{\alpha, {\bf k}, n} \, \hat{b}^\dag_{\alpha, -{\bf k}})$, in analogy with the number conserving Bogoliubov theory of weakly-interacting gases \cite{castin2001coherent}. Here, $\hat{b}_{\alpha, {\bf k}}$($\hat{b}^\dag_{\alpha, {\bf k}}$) represents the annihilation (creation) of a single collective mode in the $\alpha^\text{th}$ branch with momentum $\mathbf{k}$ and energy $\omega_{\alpha, \mathbf{k}}$, corresponding to the Bogoliubov-rotated quadratic form of the bath Hamiltonian $\hat{H}_\mathrm{B} \approx \sum_{\alpha, {\bf k}} \omega_{\alpha, \mathbf{k}} \, \hat{b}^\dag_{\alpha, \mathbf{k}} \, \hat{b}_{\alpha, \mathbf{k}}$.

The two energetically lowest excitations consist of the gapless Goldstone and gapped amplitude (Higgs) mode in the SF regime, and particle-hole excitations in the MI phase \cite{sachdev_2011, PhysRevLett.120.073201, PhysRevB.75.085106, Endres2012}. At the tip of the Mott lobe, both Goldstone and Higgs modes are gapless, and the filling is fixed and commensurate on both sides of the transition, belonging to the $O(2)$ universality class (c.f. Ref.~\cite{sachdev_2011}). Away from the lobe tip, the Goldstone mode is quadratic whereas the Higgs mode remains gapped, and the system, despite being strongly-interacting, behaves as an effective free Bose gas of quasiparticles \cite{sachdev_2011}. Due to the discontinuous change of the density, this is referred to as the CI transition. Crucially, the QGW model is able to essentially capture both universality classes of the MI/SF quantum phase transition as shown in Ref.~\cite{PhysRevResearch.2.033276, SM}.

The polaron cloud is formed by the multi-branch spectrum of bath excitations, which are taken account by expanding $\hat{H}_{IB}$ in powers of the $\hat{b}_{\alpha, {\bf k}}$ and $\hat{b}^\dag_{\alpha, {\bf k}}$ \cite{SM}, finding
\begin{equation}
    \hat{H}_{IB} \approx U_{12} \sum_{\bf r} \hat{n}_{I, {\bf r}} \left( n_0 + \delta_1 \hat{n}_{\bf r} + \delta_2 \hat{n}_{\bf r} + \dots \right) \, .
\end{equation}
The first term in parenthesis is the mean-field energy shift $U_{12} \, n_0$, where $n_0 = \sum_n n \left| c^0_n \right|^2$, while the second term reads
\begin{equation}\label{eq:d1n}
    \delta_1 \hat{n}({\bf r}) = \frac{1}{\sqrt{I}} \sum_{\alpha} \sum_{\bf k} N_{\alpha, {\bf k}} \left(\hat{b}_{\alpha, {\bf k}} \, e^{i \, {\bf k} \cdot {\bf r}} + \hat{b}^\dag_{\alpha, {\bf k}} \, e^{-i \, {\bf k} \cdot {\bf r}} \right) \, ,
\end{equation} 
with $N_{\alpha, {\bf k}} = \sum_n n \, c^0_n (u_{\alpha, {\bf k}, n} + v_{\alpha, {\bf k}, n})$. At this level, the resultant Fr\"ohlich model \cite{mahan2013many, frohlich54} is already more general than the usual Bogoliubov expansion on top of a SF state \cite{rey, PhysRevA.89.033615, cosco2018}, accurate only in the deep SF regime \cite{SM}.
However, since two-particle processes become relevant in the vicinity of the $O(2)$ critical region~\cite{caleffi2021}, the nonlinear term should also be included
\begin{align}\label{eq:d2n}
    &\delta_2 \hat{n}({\bf r}) = \frac{1}{I} \sum_{\alpha, \beta} \sum_{\bf k, p} \left[ W_{\alpha {\bf k}, \beta {\bf p}} \left( \hat{b}^\dag_{\alpha, {\bf k}} \, \hat{b}^\dag_{\beta, {\bf p}} e^{-i ({\bf k} + {\bf p}) \cdot {\bf r}} + \mathrm{h.c.} \right) \right. \nonumber \\
    &\left. + \ U_{\alpha {\bf k}, \beta {\bf p}}  \hat{b}^\dag_{\alpha, {\bf k}}  \hat{b}_{\beta, {\bf p}}  e^{i ({\bf p} - {\bf k}) \cdot {\bf r}} + V_{\alpha {\bf k}, \beta {\bf p}}  \hat{b}_{\alpha, {\bf k}}  \hat{b}^\dag_{\beta, {\bf p}}  e^{i ({\bf k} - {\bf p}) \cdot {\bf r}} \right] ,
\end{align} 
obtaining a {\it beyond}-Fr\"ohlich model of the bath-impurity interaction. At zero temperature, only contributions weighted by $W_{\alpha {\bf k}, \beta {\bf p}} = \sum_n (n - n_0) \, u_{\alpha, {\bf k}, n} \, v_{\alpha, {\bf k}, n}$ remain in the two-particle channel \cite{SM}.

\begin{figure}[!t]
\centering
\includegraphics[width=1\linewidth]{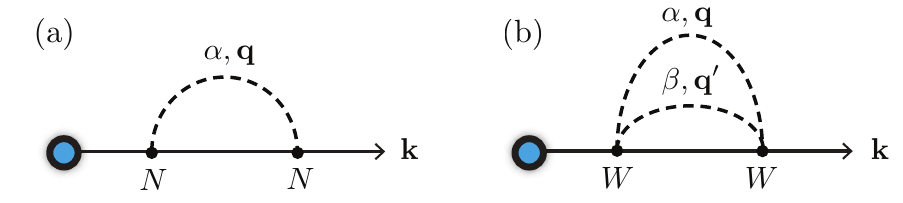}
\caption{Diagrammatic representation of the beyond mean-field contributions to the interacting impurity Green's function to second order in $U_{12}$ within the zero-temperature QGW approach. (a) and (b) depict the one- and two-particle diagrams with QGW vertex functions $N_{\alpha, {\bf q}}$ and $W_{\alpha {\bf q}, \beta {\bf q'}}$, respectively. Full lines represent the bare impurity Green's function $G^{(0)}$, while dashed lines correspond to bare Green's functions $D^{(0)}_\alpha$ of the collective modes of the BH bath.}
\label{fig:feynman}
\end{figure}

\textit{Self-energy.} --
The polaron is composed of both the impurity and the surrounding cloud of excited bath modes, producing the quasiparticle illustrated in Fig.~\ref{fig:1}(b). The dressed properties of the impurity are quantified by the self-energy $\Sigma({\bf k}, \omega) = G^{(0)}({\bf k}, \omega)^{-1} - G({\bf k}, \omega)^{-1}$, where $G^{(0)}({\bf k}, \omega) \ (G({\bf k}, \omega))$ is the bare (interacting) impurity Green's function. We calculate the self-energy diagrammatically via the Dyson series, including all relevant zero-temperature diagrams to second order in $U_{12}$ shown in Fig.~\ref{fig:feynman}. To this level of approximation within the QGW approach, we find 
\begin{align}\label{eq:selfenergy}
    \Sigma&({\bf k}, \omega) = \\ 
    &\ U_{12} \langle \hat{n} \rangle + \frac{U^2_{12}}{I} \sum_\alpha \sum_{\bf q} \frac{\left| N_{\alpha, {\bf q}} \right|^2}{\omega - \omega_{\alpha, {\bf q}} - \varepsilon_{\bf k + q} + i \, 0^+} \nonumber \\
    &+ \frac{U^2_{12}}{2 \, I^2} \sum_{\alpha, \beta} \sum_{{\bf q, q'}} \frac{\left| W_{\alpha {\bf q}, \beta {\bf q'}} + W_{\beta {\bf q'}, \alpha {\bf q}} \right|^2}{\omega - \omega_{\alpha, {\bf q}} - \omega_{\beta, {\bf q'}} - \varepsilon_{\bf k - q - q'} + i \, 0^+} \, , \nonumber
\end{align}
having one-particle vertex $N_{\alpha, {\bf k}}$ [Fig.~\ref{fig:feynman}(a)], two-particle direct $W_{\alpha {\bf k}, \beta {\bf k'}}$ and exchange $W_{\beta {\bf k'}, \alpha {\bf k}}$ vertices [Fig.~\ref{fig:feynman}(b)], and Hartree shift $U_{12} \langle \hat{n} \rangle$ where $\langle \delta_2 \hat{n}\rangle = \sum_n (n - n_0) \langle \delta \hat{c}^\dag_n({\bf r}) \, \delta \hat{c}_n({\bf r}) \rangle$. Due to the vanishing of one-particle vertices in the MI regime, we immediately notice the crucial role of two-particle processes such as the excitation of particle-hole pairs in the quantum critical regime. At zero temperature, these beyond-Fr\"ohlich effects describe only the simultaneous emission or absorption of two excitations by the impurity. This is in sharp contrast with the Fermi polaron, where the equivalent processes are suppressed due to Fermi statistics \cite{PhysRevLett.98.180402}. Furthermore, we note that Eq.~\eqref{eq:selfenergy} is invariant under sign change of $U_{12}$, up to a possible overall shift due to the Hartree contribution, and therefore we restrict to the case $U_{12} > 0$, producing results for $U_{12}/U = 0.2$ without loss of generality at weak coupling.

The key properties of the lattice Bose polaron include its dispersion, stability, and coherence, which have been previously measured for its continuum counterpart \cite{PhysRevA.85.023623, PhysRevLett.117.055301, PhysRevLett.117.055302, yan2020science, Skou2021, skou2022}.
We calculate the polaron dispersion by considering the self-energy on the mass shell $(\omega = \varepsilon_{\bf k})$. Within this Rayleigh-Schr\"odinger perturbative scheme \cite{mahan2013many}, the full polaron energy is $E_{\bf k} = \varepsilon_{\bf k} + \text{Re} \, \Sigma({\bf k}, \varepsilon_{\bf k})$, which can be expanded at low momentum as 
\begin{equation}\label{eq:polaronenergy}
    E_{\bf k} = E_0 + \frac{{\bf k}^2}{2 \, M_*} + O(k^4) \, ,
\end{equation}
where $E_0$ is the bath-induced shift of the polaron energy, and $M_*$ is the polaron effective mass. Both quantities can be inferred from Eq.~\eqref{eq:polaronenergy} via
\begin{equation}\label{eq:e0mstar}
    E_0 = \text{Re} \, \Sigma({\bf 0}, 0) \, , \quad \frac{M}{M_*} = \frac{M}{d} \sum_{i = 1}^{d} \left. \frac{\partial^2 E_{\bf k}}{\partial k^2_i}\right|_{\bf k = 0} \, .
\end{equation}
The stability and coherence of the polaron are given by the (on-shell) momentum-dependent decay rate and quasiparticle residue 
\begin{equation}\label{eq:GammaZ}
\begin{aligned}
    \Gamma(\mathbf{k}) = -2 \, \text{Im}\Sigma{\left( \mathbf{k}, \varepsilon_{\mathbf{k}} \right)},\, Z(\mathbf{k})^{-1} = 1 - \left. \frac{\partial \text{Re} \Sigma{\left( \mathbf{k}, \omega \right)}}{\partial \omega} \right|_{\omega = \varepsilon_{\mathbf{k}}},
\end{aligned}
\end{equation}
respectively. The quasiparticle residue measures the overlap between the polaron and free impurity states, quantifying the renormalized spectral weight near the polaronic pole of the Green's function \cite{mahan2013many}. The polaron is well-defined provided that $\Gamma({\bf k}) \ll E_{\bf k}$ and $Z({\bf k})$ is comparable to unity.

\begin{figure}[!t]
\centering
\includegraphics[width=1\linewidth]{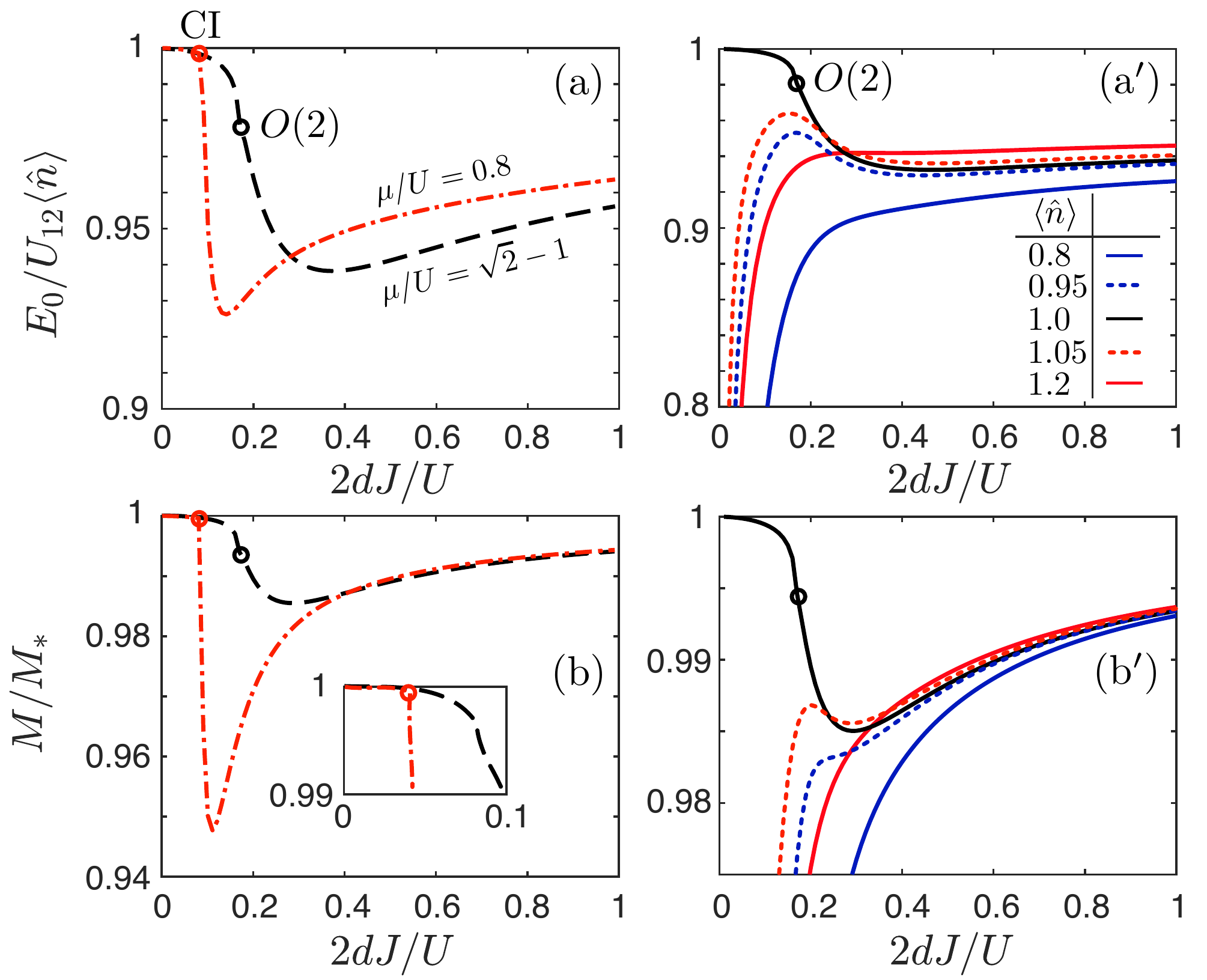}
\caption{(a)-(b) Polaron spectral properties for fixed $\mu/U$ across the $O(2)$ (black dashed) and CI (red dashed-dotted) transitions, with the non-analytic nature of the latter shown in the inset. (a$'$)-(b$'$) Polaron spectral properties for fixed $\langle \hat{n} \rangle$. The $\langle \hat{n} \rangle = 1$ line crosses the $O(2)$ critical point at $2 \, d \, J/U \approx 0.17$, while the non-integer lines approach the HC regime, where the extremal point of the CI transition is located. Dots indicate quantum critical points.}
\label{fig:2}
\end{figure}

\textit{Spectral properties.} --
We study now in detail how the spectral properties of the polaron depend on the quantum critical behavior of the bath. Within the QGW formalism, the polaron energy and effective mass are obtained from Eq.~\eqref{eq:e0mstar} \cite{SM}. Our results for $E_0$ and $M/M_*$ are shown in Fig.~\ref{fig:2} both at fixed chemical potential [(a)-(b)] and fixed filling [(a$'$)-(b$'$)] across the $O(2)$ and CI transitions. In general, starting from the deep SF regime and reducing the hopping $J/U$, the dressing effect of the bath excitations leads to a heavier polaron with energy lower than the Hartree shift, which corresponds to the limiting value in both the deep SF and MI regimes. In the former regime, the contribution of quantum fluctuations is saturated by the excitation of the gapless Goldstone mode, whereas in the latter it is solely due to the excitation of particle-hole pairs. However, as the strongly-correlated regime is approached, we observe two distinct trends depending on the universality class of the transition and the character of the underlying critical fluctuations.

Upon crossing the $O(2)$ phase transition, either via fixed $\langle n \rangle = 1$ or $\mu/U = \sqrt{2} - 1$, we see from Fig.~\ref{fig:2} that both $E_0$ and $M/M_*$ reach an absolute minimum on the SF side and increase {\it smoothly} across the transition. Here, both the Goldstone and Higgs branches make competing contributions to the polaron cloud at the one-particle level due to the closing of the Higgs gap at the critical point. For the same reason, the beyond-Fr\"ohlich two-particle process involving the coupling of the Goldstone and Higgs modes, encoded in the vertex $W_{\mathrm{G} {\bf q}, \mathrm{H} {\bf q'}}$, makes also a significant contribution on the SF side of the transition. These processes become the only non-vanishing contribution at the critical point, leading to a critical bath with non-Gaussian statistics and a smooth crossover towards mean-field values in the deep MI regime. In the MI regime, the polaron cloud is composed of doublon-holon pairs with fixed size set by $\left( J/U \right)^2$, becoming increasingly localized in a small neighborhood of the impurity with emission and absorption processes occurring on short time scales \cite{KRUTITSKY2016, PhysRevResearch.2.033276, caleffi2021}.

The situation changes drastically when the MI boundary is crossed instead at the CI transition for fixed chemical potential, as shown in Fig.~\ref{fig:2}(a)-(b) [red dashed-dotted line]. In this case, both $E_0$ and $M/M_*$ are dominated by one-particle processes despite strong interactions. These are predominantly due to the Goldstone mode, as the Higgs mode remains gapped at the transition. On the SF side of the CI point, the polaron properties are more strongly renormalized than at the $O(2)$ point, moving {\it sharply} towards their bare values on the MI side. At the CI point, the bath behaves as an effective free Bose gas of quasiparticles with vanishing sound speed and large compressibility, hence becoming softer to perturbations of the density than at the $O(2)$ point where the sound speed remains finite and the compressibility vanishes \cite{PhysRevA.84.033602, PhysRevResearch.2.033276}. Consequently, there is a stronger interplay between density fluctuations of the bath and the impurity. The non-analyticity of the polaron properties at the transition [inset Fig.~\ref{fig:2}(b)] closely reflects the discontinuous behavior of the single-particle coherence length for the CI transition \cite{sachdev_2011, PhysRevResearch.2.033276} due to the abrupt suppression of one-particle processes, with only two-particle processes remaining.

Traveling instead along lines of non-integer filling [see Fig.~\ref{fig:1}(a)], the bath enters the regime of the so-called hard-core (HC) SF state ($J/U \ll 1$) at the borders of the Mott lobes shown in Fig.~\ref{fig:2}(a$'$)-(b$'$) [red and blue solid and dotted lines]. Here, the bath becomes strongly-interacting without entering the MI phase, with the polaron cloud again dominated by one-particle excitations of the Goldstone mode having almost quadratic dispersion. The behavior of the hard-core SF as a free Bose gas of strongly-renormalized quasiparticles entails a diverging compressibility and, correspondingly, a divergence in $E_0$ and $M/M_*$ precisely at the HC point $J/U \to 0$. As the strongly-correlated regime is approached, a non-monotonic behavior is found as the filling nears an integer value [dotted lines Fig.~\ref{fig:2}(a$'$)-(b$'$)] due to the increased vicinity to the phase transition.

\begin{figure}[!t]
\centering
\includegraphics[width=1\linewidth]{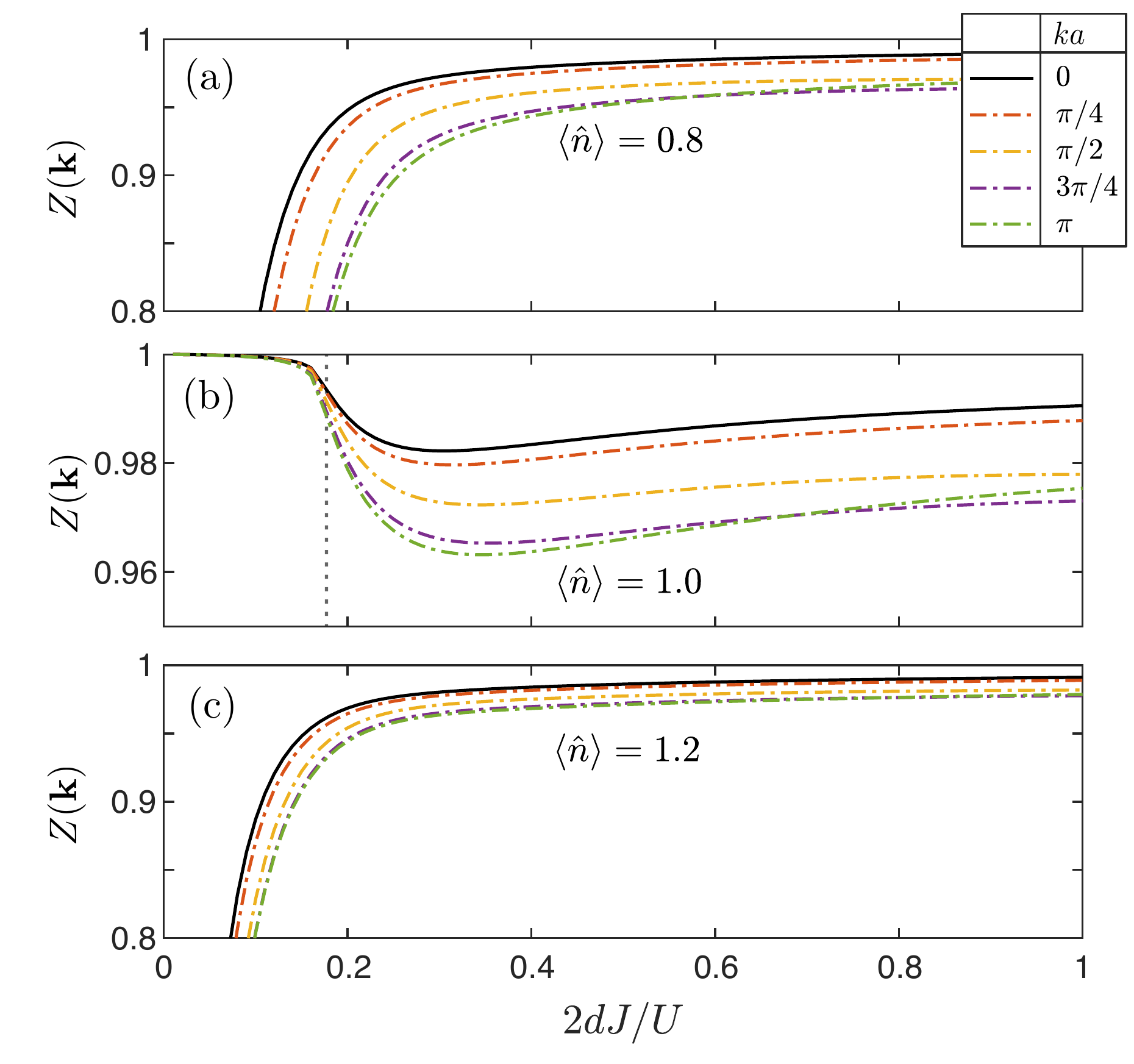}
\caption{Momentum dependence of the quasiparticle residue $Z(\mathbf{k})$ for fixed density (a) $\langle \hat{n} \rangle = 0.8$, (b) $\langle \hat{n} \rangle=1.0$, and (c) $\langle \hat{n} \rangle = 1.2$. Whereas the polaron is well defined ($Z \sim 1$) across the $O(2)$ transition (dotted vertical line), it suffers from an orthogonality catastrophe ($Z \to 0$) in the HC regime along lines of non-integer filling.} 
\label{fig:3}
\end{figure}

\textit{Coherence properties.} --
The quantum critical nature of the bath can also strongly impact on the coherence and stability of the polaron, posing limitations on its experimental detection. The on-shell quasiparticle residue and decay rate follow from Eqs.~\eqref{eq:GammaZ}, with the latter resulting in a Fermi's Golden Rule with effective couplings set by the one- and two-particle vertices \cite{SM}. The decay rate mirrors the zero temperature spontaneous emission of excitations from the polaron cloud, requiring $\varepsilon_{\bf k} > \omega_{\alpha, {\bf k}}$. Physically these processes become increasingly likely for lighter impurities due to the concavity of $\varepsilon_{\bf k}$.

Our results for $Z({\bf k})$ are shown in Fig.~\ref{fig:3} at fixed filling across the $O(2)$ transition and towards the HC points. As the bath crosses the $O(2)$ transition at fixed filling $\langle \hat{n} \rangle = 1$ [Fig.~\ref{fig:3}(b)], the polaron remains well-defined ($Z \sim 1$) at all momenta. However, along lines of non-integer filling [Fig.~\ref{fig:3}(a) and (c)], the bath enters the regime of the hard-core SF, which has been recently predicted to spoil the coherence of static impurities over time \cite{caleffi2021}. In this region, it is expected that the motion of a mobile impurity is diffusive due to the almost free-particle nature of the bath \cite{Lampo2017bose}. The time-dependent description of this phenomenon is left for future work. Eventually, the mobile impurity is so strongly renormalized that it becomes orthogonal with its bare state $(Z \to 0)$, giving rise to a bosonic instance of Anderson's orthogonality catastrophe \cite{PhysRevLett.18.1049}. This effect occurs also in the continuum for a mobile impurity immersed in an ideal Bose gas due to the infinite compressibility of the bath and macroscopic polaron cloud that forms \cite{PhysRevA.103.013317, PhysRevX.8.011024, Sun2004}. The behavior of the mobile impurity at fixed chemical potential crossing the CI transition is intermediate to the limiting cases of incoherent [HC point] and coherent [$O(2)$ point] character. Along the line of CI transitions, $Z({\bf k})$ displays an increasingly sharp transition analogous to Fig.~\ref{fig:1}(a$'$)-(b$'$) at all momenta, approaching the orthogonality catastrophe at the HC point \cite{SM}.

Across the $O(2)$ transition, either for fixed filling or chemical potential, we find that the polaronic quasiparticle is generally long-lived ($\Gamma \sim 0$). However, in the vicinity of the orthogonality catastrophe, the combination of polaron incoherence and expected diffusive nature of the bath leads to $\Gamma({\bf k}) \gtrsim E_{\bf k}$, such that the quasiparticle picture is lost. In fact, the criterion $\varepsilon_{\bf k} > \omega_{\alpha, {\bf k}}$ for spontaneous emission is aided at low momentum by the strongly-renormalized free particle dispersion of the bath excitations. From the behavior of $E_0$ and $M/M_*$ shown in Fig.~\ref{fig:2}(a)-(b), it is clear that the eventual acquisition of negative values signals an obvious breakdown of the assumption that the back action of the impurity on the bath can be neglected, as the environment becomes energetically and thermodynamically unstable to the presence of the impurity.

{\it Conclusion.} --
By blending diagrammatic techniques with the recently developed quantum Gutzwiller method, we have presented a beyond-Fr\"ohlich study of the physics of a mobile impurity confined in a lattice in a Bose-Hubbard bath throughout the entire phase diagram. Crucially, we have shown how the different universality classes of phase transitions occurring within the bath are strongly reflected in the properties of the polaron via the cloud of collective modes that dress the impurity. This finding highlights the experimental potential of polarons as versatile probes of quantum correlations within a many-body environment. The success of the approach introduced in this Letter raises exciting prospects for investigating impurity physics on lattice systems of interest within the larger context of quantum simulation \cite{Gross2017, RevModPhys.86.153}.


\begin{acknowledgments}
\textit{Acknowledgments.} -- We thank Georg M. Bruun for fruitful discussions. This project has received financial support from Provincia Autonoma di Trento and the Italian MIUR through the PRIN2017 project CEnTraL (Protocol Number 20172H2SC4).
\end{acknowledgments}

\bibliographystyle{apsrev4-1}
\bibliography{biblio}


\end{document}


\title{Supplementary Material \texorpdfstring{\\}{} ``Lattice polarons across the superfluid to Mott insulator transition"} 
\author{V. E. Colussi}
\email[Corresponding author: ]{colussiv@gmail.com}
\affiliation{INO-CNR BEC Center and Dipartimento di Fisica, Universit\`a di Trento, Via Sommarive 14, 38123 Povo, Trento, Italy}
\author{F. Caleffi}
\affiliation{International School for Advanced Studies (SISSA), Via Bonomea 265, I-34136 Trieste, Italy}
\author{C. Menotti}
\affiliation{INO-CNR BEC Center and Dipartimento di Fisica, Universit\`a di Trento, Via Sommarive 14, 38123 Povo, Trento, Italy}
\author{A. Recati}
\affiliation{INO-CNR BEC Center and Dipartimento di Fisica, Universit\`a di Trento, Via Sommarive 14, 38123 Povo, Trento, Italy}
\affiliation{Trento Institute for Fundamental Physics and Applications, INFN, Via Sommarive 14, 38123 Povo, Trento, Italy}

\date{\today}
\maketitle

\section{Quantum Gutzwiller Approach}\label{sm_sec:qgw}
\subsection{The QGW method in a nutshell}\label{sm_subsec:qgw_method}
The QGW approach combines the successful features of the Gutzwiller approximation \cite{gutzwiller_1991} and the Bogoliubov theory of weakly-interacting gases \cite{rey} in order to develop a robust quantum many-body theory of a generic interacting lattice model. Building on the solution of the time-dependent Gutzwiller approximation \cite{PhysRevA.84.033602}, fluctuations on top of the mean-field ground state are quantized in terms of the elementary many-body excitations of the system and systematically included in the calculation of ground state expectation values. In spite of the local nature of the underlying Gutzwiller ansatz -- see Eq.~\eqref{Gutzwiller_ansatz} below --, the QGW approach accurately reproduce both local and non-local correlations across the different phases of the BH model with minimal numerical effort.
Let us also mention that the QGW, when only quadratic fluctuations are considered, has a number of features in common with including quantum fluctuations by slave boson approaches (see in particular \cite{roscilde}, where the slave boson approach has been applied to the BH Hamiltonian to determine its entanglement entropy along its phase diagram).

Following the main derivation steps of \cite{PhysRevResearch.2.033276}, we briefly review the essential features of QGW technique, that we employ for a systematic evaluation of quantum correlations in the BH bath.

Our starting point is the Gutzwiller ansatz
\begin{equation}\label{Gutzwiller_ansatz}
    \left| \Psi_G \right\rangle = \bigotimes_{\mathbf{r}} \sum_n c_n{\left( \mathbf{r} \right)} \left| n, \mathbf{r} \right\rangle \, ,
\end{equation}
where the wave function is site-factorized
and the complex amplitudes $c_n{\left( \mathbf{r} \right)}$ of each local Fock state $\left| n, \mathbf{r} \right\rangle$ are variational parameters with normalization condition $\sum_n \left| c_n{\left( \mathbf{r} \right)} \right|^2 = 1$. Drawing on the the formal simplicity of \eqref{Gutzwiller_ansatz}, we can reformulate the BH model in terms of the following Lagrangian functional
\begin{equation}\label{lagrangian}
\begin{aligned}
    \mathfrak{L}{\left[ c, c^* \right]} &= \big\langle \Psi_G \big| \, i \, \hbar \, \partial_t - \hat{H}_B \, \big| \Psi_G \big\rangle \\
    &= \frac{i \, \hbar}{2} \sum_{\mathbf{r},n} [c^{*}_n(\mathbf{r}) \dot{c}_n(\mathbf{r}) - \textrm{c.c.}] + J \sum_{\langle \mathbf{r}, \mathbf{s} \rangle} \left[ \psi^{*}{\left( \mathbf{r} \right)} \, \psi{\left( \mathbf{s} \right)} + \text{c.c.} \right] - \sum_{\mathbf{r},n} H_n \left| c_n{\left( \mathbf{r} \right)} \right|^2 \, .
\end{aligned}
\end{equation}
In the previous equation, the dot indicates the temporal derivative,
\begin{equation}
    H_n = \frac{U}{2} n \left( n - 1 \right) - \mu \, n
\end{equation}
are the matrix elements of the on-site terms of the BH Hamiltonian $\hat{H}_B$ in Fock space and
\begin{equation}\label{order_parameter}
    \psi{\left( \mathbf{r} \right)} = \big\langle \hat{a}_{\mathbf{r}} \big\rangle = \sum_n \sqrt{n} \, c^*_{n - 1}{\left( \mathbf{r} \right)} \, c_n{\left( \mathbf{r} \right)}
\end{equation}
is the mean-field order parameter. Within this formulation, the conjugate momenta of the parameters $c_n{\left( \mathbf{r} \right)}$ are $c^*_n{\left( \mathbf{r} \right)} = \partial \mathfrak{L}/\partial \dot{c}_n{\left( \mathbf{r} \right)}$. The classical Euler-Lagrange equations associated to Lagrangian \eqref{lagrangian} are the so-called \textit{time-dependent Gutzwiller equations} as derived, e.g., in \cite{sheshandri_krishnamurthy_pandit_ramakrishnan, PhysRevA.84.033602}. In a uniform system, the stationary solutions are homogeneous: in particular, the system is found in the MI state if $\psi{\left( \mathbf{r} \right)} = 0$ and in the SF phase otherwise.

In order to go beyond the Gutzwiller approximation introduced above, it is natural to consider how quantum effects populate the excitation modes of the system and to investigate how they affect the observable quantities. We include quantum fluctuations by building a theory of the excitations starting from Lagrangian \eqref{lagrangian} via canonical quantization \cite{cohen, Ripka1985}, namely promoting the coordinates of the theory and their conjugate momenta to operators and imposing equal-time canonical commutation relations
\begin{equation}
    \left[ \hat{c}_n{\left( \mathbf{r} \right)}, \hat{c}^{\dagger}_m{\left( \mathbf{s} \right)} \right] = \delta_{\mathbf{r},\mathbf{s}} \, \delta_{n,m} \, .
\end{equation}
In analogy with the Bogoliubov approximation for dilute Bose-Einstein condensates \cite{pitaevskii2016bose, castin2001coherent}, we expand the operators $\hat{c}_n$ around their ground state values $c^0_n$, obtained by minimizing the energy $\big\langle \Psi_G \big| \hat{H}_B \big| \Psi_G \big\rangle$, as 
\begin{equation}
    \hat{c}_n{\left( \mathbf{r} \right)} = \hat{A}{\left( \mathbf{r} \right)} \, c^0_n + \delta \hat{c}_n{\left( \mathbf{r} \right)} \, .
\end{equation}
The \textit{normalization operator} $\hat{A}{\left( \mathbf{r} \right)}$ is a functional of $\delta \hat{c}_n\left( \mathbf{r} \right)$ and $\delta \hat{c}^{\dagger}_n\left( \mathbf{r} \right)$ and ensures the proper normalization $\sum_n \hat{c}^{\dagger}_n{\left( \mathbf{r} \right)} \, \hat{c}_n{\left( \mathbf{r}\right)} = \hat{\mathds{1}}$. By restricting to local fluctuations orthogonal to the ground state $\sum_n \delta \hat{c}^\dagger_n{\left( \mathbf{r} \right)} \, c^{0}_n = 0$ one has
\begin{equation}
    \hat{A}{\left( \mathbf{r} \right)} = \left[ 1 - \sum_n \delta \hat{c}^{\dagger}_n{\left( \mathbf{r} \right)} \, \delta \hat{c}_n{\left( \mathbf{r}\right)} \right]^{1/2} \, .
\end{equation}
In a homogeneous system, it is convenient to work in momentum space by writing
\begin{equation}\label{c_FT}
    \delta \hat{c}_n{\left( \mathbf{r} \right)} \equiv I^{-1/2} \sum_{\mathbf{k} \in \text{BZ}} e^{i \mathbf{k} \cdot \mathbf{r}} \, \delta \hat{C}_n{\left( \mathbf{k} \right)} \, .
\end{equation}
Inserting Eq.~\eqref{c_FT} in $\left\langle \Psi_G \right| \hat{H}_{BH} \left| \Psi_G \right\rangle$ and keeping only terms up to the quadratic order in the fluctuations, we obtain
\begin{equation}\label{H_QGW_unrotated}
    \hat{H}^{\left( 2 \right)}_{QGW} = \frac{1}{2} \mathlarger\sum_{\mathbf{k}}
    \left[ \delta \underline{\hat{C}}^{\dagger}{\left( \mathbf{k} \right)} , -\delta \underline{\hat{C}}{\left( -\mathbf{k} \right)} \right] \hat{\mathcal{L}}_{\mathbf{k}}
    \begin{bmatrix}
    \delta \underline{\hat{C}}{\left( \mathbf{k} \right)} \\
    \delta \underline{\hat{C}}^{\dagger}{\left( -\mathbf{k} \right)}
    \end{bmatrix} \, ,
\end{equation}
up to a constant energy shift equal to the BH ground state energy. Here, the vector $\delta \underline{\hat{C}}(\mathbf{k})$ gathers the components $\delta{\hat{C}}_n(\mathbf{k})$, and $\hat{\mathcal{L}}_{\mathbf{k}}$ is a pseudo-Hermitian matrix, for the explicit expression of which we refer the interested reader to \cite{PhysRevResearch.2.033276}. A suitable Bogoliubov rotation of the Gutzwiller operators in terms of the fundamental excitation modes of the system
\begin{equation}\label{c_rotation}
    \delta \hat{C}_n{\left( \mathbf{k} \right)} = \sum_{\alpha} u_{\alpha, \mathbf{k}, n} \, \hat{b}_{\alpha, \mathbf{k}} + \sum_{\alpha} v^{*}_{\alpha, -\mathbf{k}, n} \, \hat{b}^{\dagger}_{\alpha, -\mathbf{k}} \, ,
\end{equation}
recasts the quadratic form \eqref{H_QGW_unrotated} into the desired diagonal form
\begin{equation}\label{H_QGW}
    \hat{H}_B \approx \sum_{\alpha} \sum_{\mathbf{k}} \omega_{\alpha, \mathbf{k}} \, \hat{b}^{\dagger}_{\alpha, \mathbf{k}} \, \hat{b}_{\alpha, \mathbf{k}} \, ,
\end{equation}
where each $\hat{b}_{\alpha, \mathbf{k}}$ corresponds to a different many-body excitation mode with frequency $\omega_{\alpha,\mathbf{k}}$, labeled by its momentum $\mathbf{k}$ and branch index $\alpha$. Bosonic commutation relations between the annihilation and creation operators $\hat{b}_{\alpha, \mathbf{k}}$ and $\hat{b}^\dagger_{\alpha, \mathbf{k}}$,
\begin{equation}
    \left[ \hat{b}_{\alpha, \mathbf{k}}, \hat{b}^{\dagger}_{\alpha', \mathbf{k'}} \right] = \delta_{\mathbf{k}, \mathbf{k}'} \, \delta_{\alpha, \alpha'} \, ,
\end{equation}
are enforced by choosing the usual Bogoliubov normalization condition 
\begin{equation}
    \underline{u}^{*}_{\alpha, \mathbf{k}} \cdot \underline{u}_{\beta, \mathbf{k}} - \underline{v}^{*}_{\alpha, -{\mathbf{k}}} \cdot \underline{v}_{\beta, -\mathbf{k}} = \delta_{\alpha \beta} \, ,
\end{equation}
where the vectors $\underline{u}_{\alpha, \mathbf{k}}$ ($\underline{v}_{\alpha, \mathbf{k}}$) contain the components $u_{\alpha, \mathbf{k}, n}$ ($v_{\alpha, \mathbf{k}, n}$).

\subsection{QGW quantization protocol for observables}\label{sm_subsec:protocol}
The effective, quadratic description of the BH environment in terms of its collective modes \eqref{H_QGW} provided by the QGW approach entails a straightforward scheme for expressing the original bath operators in terms of the bath excitations. Based on the quantization procedure outlined before, the decomposition of any observable $\hat{O}{\left( \hat{a}^{\dagger}_{\mathbf{r}}, \hat{a}_{\mathbf{r}} \right)}$ along the basis of the collective modes consists in applying a four-step procedure that we summarize as follows:
\begin{enumerate}
    \item Determine the expression $\mathcal{O}{\left[ c, c^* \right]} = \big\langle \Psi_G \big| \hat{O} \big| \Psi_G \big\rangle$ in terms of the Gutzwiller parameters $c_n$ and $c^*_n$;
    \item Create the operator $\hat{\mathcal{O}}{\left[ \hat{c}, \hat{c}^{\dagger} \right]}$ by replacing the Gutzwiller parameters in $\mathcal{O}{\left[ c, c^* \right]}$ by the corresponding operators $\hat{c}_n{\left( \mathbf{r} \right)}$ and $\hat{c}^{\dagger}_n{\left( \mathbf{r} \right)}$ without modifying their ordering;
    \item Expand the operator $\hat{\mathcal{O}}$ order by order in the fluctuations $\delta \hat{c}_n$ and $\delta \hat{c}^{\dagger}_n$, taking into account the dependence of the operator $\hat{A}$ on the fluctuation operators. The contribution of $\hat{A}$ may be of fundamental importance when higher orders in the fluctuations become relevant, e.g. in the strongly-correlated regime of the system;
    \item Taking advantage of the quadratic character of the QGW Hamiltonian, invoke Wick theorem to compute the expectation value of products of operators on Gaussian states -- such as the ground or a thermal state obtained from $\hat{H}_B$.
\end{enumerate}

For the purpose of the present work, we are interested in the coupling of the impurity with the bath modes excited in the density channel, according to our microscopic model (1). Applying the QGW quantization protocol to the local density operator, the local density operator maps into
\begin{equation}\label{eq:density_operator}
    \hat{n}({\bf r}) = \sum_n n \, \hat{c}^\dagger_n({\bf r}) \, \hat{c}_n({\bf r}) \, .
\end{equation}
Expanding the $\hat{c}$'s to lowest order in the fluctuations, one finds
\begin{equation}
    \hat{n}({\bf r}) \approx n_0 + \delta_1 \hat{n}({\bf r}) = n_0 + \sum_n n \, c^0_n \left[ \delta \hat{c}_n({\bf r}) + \delta \hat{c}^\dagger_n({\bf r}) \right] \, ,
\end{equation}
where $n_0 = \sum_n n \left| c^0_n \right|^2$ is the mean-field density of the BH environment. Using Eqs.~\eqref{c_FT}-\eqref{c_rotation}, the first-order operator $\delta_1 \hat{n}({\bf r})$ can be readily unfolded in terms of the quantized BH modes as
\begin{equation}
    \delta_1 \hat{n}({\bf r}) = \frac{1}{\sqrt{I}} \mathlarger\sum_{\alpha} \mathlarger\sum_{\bf k} N_{\alpha, {\bf k}} \left( e^{i \, {\bf k} \cdot {\bf r}} \, \hat{b}_{\alpha, {\bf k}} + e^{-i \, {\bf k} \cdot {\bf r}} \, \hat{b}^\dagger_{\alpha, {\bf k}} \right) \, ,
\end{equation}
from which we recover the Fr\"olich-type coupling of Eq.~(3). The second-order expansion of the density operator \eqref{eq:density_operator} has to be performed more carefully, as it involves the inclusion of those fluctuation terms introduced by the normalization operator $\hat{A}{\left( \mathbf{r} \right)}$. In particular, we obtain
\begin{equation}\label{eq:d2n}
    \delta_2 \hat{n}({\bf r}) = \sum_n n \, \delta \hat{c}^\dagger_n({\bf r}) \, \delta \hat{c}_n({\bf r}) - n_0 \, \hat{F} \, ,
\end{equation}
where we have defined $\hat{F} = 1 - \hat{A}^2({\bf r})$. 

It is interesting to observe that the second-order quantum correction to local density field is given by the sum of two distinct terms, one given by quantum fluctuations only and the other, proportional to the mean-field average, deriving exclusively from the normalization operator via the operator $\hat{F}$, which works as a control parameter of the theory. This result makes more explicit the physical role of $\hat{A}{\left( {\bf r} \right)}$, which accounts for the feedback of quantum fluctuations onto the Gutzwiller mean-field state.

Finally, we can rephrase Eq.~\eqref{eq:d2n} in terms of two-body excitations of the collective modes as
\begin{equation}\label{eq:d2n_two_body}
    \delta_2 \hat{n}({\bf r}) = \frac{1}{I} \sum_{\alpha, \beta} \sum_{\bf k, p} \left[ W_{\alpha {\bf k}, \beta {\bf p}} \left( \hat{b}^\dagger_{\alpha, {\bf k}} \, \hat{b}^\dagger_{\beta, {\bf p}} e^{-i ({\bf k} + {\bf p}) \cdot {\bf r}} + \mathrm{h.c.} \right) + U_{\alpha {\bf k}, \beta {\bf p}} \, \hat{b}^\dagger_{\alpha, {\bf k}} \, \hat{b}_{\beta, {\bf p}} \, e^{i ({\bf p} - {\bf k}) \cdot {\bf r}} + V_{\alpha {\bf k}, \beta {\bf p}} \, \hat{b}_{\alpha, {\bf k}} \, \hat{b}^\dagger_{\beta, {\bf p}} \, e^{i ({\bf k} - {\bf p}) \cdot {\bf r}} \right] \, ,
\end{equation}
where we have introduced the two-mode vertex factors
\begin{equation}\label{density_vertices}
\begin{gathered}
W_{\alpha \mathbf{k}, \beta \mathbf{p}} = \sum_n \left( n - n_0 \right) u_{\alpha, \mathbf{k}, n} \, v_{\beta, \mathbf{p}, n} \, , \\
U_{\alpha \mathbf{k}, \beta \mathbf{p}} = \sum_n \left( n - n_0 \right) u_{\alpha, \mathbf{k}, n} \, u_{\beta, \mathbf{p}, n} \, , \\
V_{\alpha \mathbf{k}, \beta \mathbf{p}} = \sum_n \left( n - n_0 \right) v_{\alpha, \mathbf{k}, n} \, v_{\beta, \mathbf{p}, n} \, ,
\end{gathered}
\end{equation}
whose derivation is extensively discussed in \cite{PhysRevResearch.2.033276}. Importantly, the operator \eqref{eq:d2n_two_body} underlies the beyond-Fr\"olich physics explored in the present work.

For the sake of completeness, we mention here that our results for the beyond-Fr\"olich effects originated by the second-order expansion \eqref{eq:d2n} have been restricted to sufficiently strong interactions $\left( 2dJ/U \lesssim 1 \right)$ because of well-known issues regarding the overestimation of quantum fluctuations in a $d = 2$ bosonic system in the weakly-interacting limit, which would require a careful renormalization of the two-body scattering length, see in particular \cite{SciPostPhys.12.3.111} for a more detailed discussion.

We conclude this section by reporting for later convenience the first-order expansion of the order parameter field \eqref{order_parameter} provided by our quantization scheme,
\begin{equation}
\hat{\psi}({\bf r}) = \sum_n \sqrt{n} \, \hat{c}^\dagger_{n - 1}({\bf r}) \, \hat{c}_n({\bf r}) \approx \psi_0 + \delta_1 \hat{\psi}({\bf r}) \, ,
\end{equation}
where $\psi_0$ corresponds to the mean-field condensate density and
\begin{equation}\label{eq:QGW_psi}
\delta_1 \hat{\psi}({\bf r}) = \frac{1}{\sqrt{I}} \mathlarger\sum_{\alpha} \mathlarger\sum_{\bf k} \left[ U_{\alpha, {\bf k}} \, e^{i \, {\bf k} \cdot {\bf r}} \, \hat{b}_{\alpha, {\bf k}} + V_{\alpha, {\bf k}} \, e^{-i \, {\bf k} \cdot {\bf r}} \, \hat{b}^\dagger_{\alpha, {\bf k}} \right] \, .
\end{equation}
This result clearly shows that the spectral amplitudes $U_{i, \alpha, {\bf k}}$ $\left( V_{i, \alpha, {\bf k}} \right)$ quantify the particle (hole) character of the excitation $\left( \alpha, {\bf k} \right)$ for the $i^\text{th}$ in the one-body channel. Indeed, we remark that our approximation for $\hat{\psi}({\bf r})$ plays the genuine role of Bose field operator in the BH system, as it can be shown to satisfy bosonic commutation relations exactly \cite{PhysRevResearch.2.033276, SciPostPhys.12.3.111}. Moreover, we mention that $U_{i, \alpha, {\bf k}}$ $\left( V_{i, \alpha, {\bf k}} \right)$ converge exactly to their counterparts within Bogoliubov's theory in the deep SF regime,
\begin{equation}\label{eq:bogoliubov_amplitudes}
\left| U_{\bf k} \right|^2 = \frac{1}{2} \left[ \frac{\varepsilon({\bf k}) + \left| \psi_0 \right|^2 U}{\omega_{\bf k}} + 1 \right] \, , \qquad \left| V_{\bf k} \right|^2 = \frac{1}{2} \left[ \frac{\varepsilon({\bf k}) + \left| \psi_0 \right|^2 U}{\omega_{\bf k}} - 1 \right] \, ,
\end{equation}
where
\begin{equation}
\omega_{\bf k} = \sqrt{\varepsilon({\bf k}) \left[ \varepsilon({\bf k}) + 2 \left| \psi_0 \right|^2 U \right]}
\end{equation}
is the usual Goldstone mode dispersion of a weakly-interacting Bose gas.

\subsection{Fr\"ohlich models}\label{sm_subsec:frohlich_models}
In this section, we analyze the polaron problem in the limit where the quantum depletion of Bose-Einstein condensate (BEC) forming in the SF regime is small compared to the lattice filling $\langle \hat{n} \rangle$. In this case, the density of the cloud of excitations surrounding the impurity is expected to be small relative to the density of the surrounding BEC, justifying the usual Bogoliubov expansion in powers of the BEC density $\psi_0$. Therefore, we can choose to expand (and truncate) the Hamiltonian (1) in terms of BEC fluctuations (rather than in the density channel) to obtain the Bogoliubov form of the Fr\"ohlich Hamiltonian routinely used in the characterization of a mobile impurity in a weakly-interacting BEC (c.f.~\cite{grusdt2015new}).

Within the usual Bogoliubov approximation, one has just a single mode, the Goldstone phonon excitation, such that the bath-impurity interaction takes the form \cite{PhysRevA.89.033615}
\begin{equation}\label{eq:hibbogo}
\hat{H}_\mathrm{IB} \approx U_{12} \left| \psi_0 \right|^2 + \frac{U_{12}}{\sqrt{I}} \sum_{\bf k} B_{\bf k} \, e^{i \, {\bf k} \cdot {\bf r}} \left( \hat{b}_{\bf k} + \hat{b}^\dagger_{-{\bf k}} \right)
\end{equation}
where
\begin{equation}
B_{\bf k} = \psi_0 \left( U_{\bf k} + V_{\bf k} \right)
\end{equation}
is the one-particle vertex function of the Goldstone mode, with the particle (hole) amplitudes $U_{\bf k} \left( V_{\bf k} \right)$ given by the analytic expressions in Eq.~\eqref{eq:bogoliubov_amplitudes}. Following the terminology of \cite{grusdt2015new}, we refer to the Hamiltonian \eqref{eq:hibbogo} for the Bose polaron interaction as the \textit{Bogoliubov-Fr\"ohlich model}.

The above simple model can be generalized to include the interaction of the impurity with all the other excitations of the background bath, e.g. the amplitude (Higgs) mode in the SF state and doublon-holon modes in the MI regime. This simply amounts to replace the standard Bogoliubov expansion of the Bose field operators with the QGW lowest-order projection \eqref{eq:QGW_psi}, which is nothing but a direct generalization of the former to comprise the additional excitation modes $\alpha$ that become important away from the weakly-interacting limit. In what follows, we refer to this extended representation as the \textit{QGW Bogoliubov-Fr\"ohlich model} of the BH polaron, specified by the multi-branch vertex functions $B_{\alpha, \bf k} = \psi_0 \left( U_{\alpha, \bf k} + V_{\alpha, \bf k} \right)$ weighting the coupling of the impurity with one-body condensate excitations across the whole phase diagram of the bath.

\section{Polaron Self-Energy and Related Properties}\label{sm_sec:self_energy}
We calculate the self-energy of the polaron diagrammatically via Dyson's equation for the interacting impurity Green's function (c.f. \cite{mahan2013many})
\begin{equation}
G({\bf k}, \omega) = \frac{G^{(0)}({\bf k}, \omega)}{1 - G^{(0)}({\bf k}, \omega) \, \Sigma({\bf k}, \omega)} \, ,
\end{equation}
evaluated at zero temperature. The total self-energy $\Sigma({\bf k}, \omega)$ is obtained by virtually summing the infinite number of irreducible self-energy diagrams, however we consider only contributions up to second order in the coupling strength $U_{12}$, approximating
\begin{equation}\label{eq:dysonapprox}
    \Sigma({\bf k}, \omega) \approx U_{12} \langle \hat{n} \rangle + \Sigma_\mathrm{1P}({\bf k}, \omega) + \Sigma_\mathrm{2P}({\bf k}, \omega) \, ,
\end{equation}
where the one-particle (Gutzwiller-Fr\"ohlich) contribution is
\begin{equation}
    \Sigma_\mathrm{1P}({\bf k}, \omega) = \frac{U^2_{12}}{I} \mathlarger{\sum_\alpha \sum_{\bf q}} \frac{\left| N_{\alpha, {\bf q}} \right|^2}{\omega - \omega_{\alpha, {\bf q}} - \varepsilon_{\bf k + q} + i \, 0^+} \, ,
\end{equation}
and the two-particle (beyond-Fr\"ohlich) contribution is 
\begin{equation}
    \Sigma_\mathrm{1P}({\bf k}, \omega) = \frac{U^2_{12}}{2 \, I^2} \mathlarger{\sum_{\alpha, \beta} \sum_{{\bf q, q'}}} \frac{\left| W_{\alpha {\bf q}, \beta {\bf q'}} + W_{\beta {\bf q'}, \alpha {\bf q}} \right|^2}{\omega - \omega_{\alpha, {\bf q}} - \omega_{\beta, {\bf q'}} - \varepsilon_{\bf k - q - q'} + i \, 0^+} \, .
\end{equation}
We note that additional two-particle processes with vertices given by $U_{\alpha{\bf k}, \beta{\bf p}}$ and $V_{\alpha{\bf k}, \beta{\bf p}}$ will contribute only at finite temperature. Additionally, we note that results for the relative behaviors of the mean-field and quantum fluctuation contributions to $\langle \hat{n} \rangle = n_0 + \langle \delta_2 \hat{n}({\bf r}) \rangle$ have been discussed in detail elsewhere \cite{PhysRevResearch.2.033276}.

For the purpose of comparison, we also report the expression for the polaron self-energy within the Bogoliubov-Fr\"ohlich model discussed in Sec.~\ref{sm_subsec:frohlich_models}
\begin{equation}
\Sigma({\bf k}, \omega) = U_{12} \left|\psi_0 \right|^2 + \frac{U_{12}^2}{I} \mathlarger{\sum_\alpha \sum_{\bf q}} \frac{\left| B_{\alpha, {\bf q}} \right|^2}{\omega - \omega_{\alpha, {\bf q}} - \varepsilon_{\bf k + q} + i \, 0^+} \, ,
\end{equation}
where, the summation is taken over the Goldstone mode only in the case of standard Bogoliubov's theory and over the entire BH multi-branch spectrum in the case of the QGW theory.

\subsection{QGW expressions for polaron properties}\label{sm_subsec:explicit_results}
Experimentally relevant properties of the polaron can be extracted from its self-energy as discussed in the main text. Here we give their explicit expressions within the QGW formalism. The quantities $E_0$, $M_*$, and $Z({\bf k})$ can be obtained straightforwardly by evaluating the real part of the Eq.~\eqref{eq:dysonapprox}. For the polaron energy we obtain
\begin{align}
    E_0 &= \text{Re} \, \Sigma({\bf 0}, 0) \\
    &= \ U_{12} \langle \hat{n} \rangle - \frac{U^2_{12}}{I} \mathlarger{\sum_\alpha \sum_{\bf q}} \frac{\left| N_{\alpha, {\bf q}} \right|^2}{\omega_{\alpha,{\bf q}} + \varepsilon_{\bf q}} - \frac{U^2_{12}}{2 \, I^2} \mathlarger{\sum_{\alpha, \beta} \sum_{{\bf q, q'}}} \frac{\left| W_{\alpha {\bf q}, \beta {\bf q'}} + W_{\beta{\bf q'}, \alpha{\bf q}} \right|^2}{\omega_{\alpha, {\bf q}} + \omega_{\beta, {\bf q'}} + \varepsilon_{\bf q + q'}} \, .
\end{align}
For the effective mass we get
\begin{align}
    \frac{M}{M_*} &= \frac{M}{d} \sum_{i = 1}^{d} \left. \frac{\partial^2 E_{\bf k}}{\partial k^2_i}\right|_{\bf k = 0} \\
    &= 1 - \frac{1}{d} \frac{U^2_{12}}{I} \mathlarger{\sum_\alpha \sum_{\bf q} \sum_{i = 1}^{d}} \left| N_{\alpha, {\bf q}} \right|^2 \left[\frac{4 \, J \, \sin(q_i \, a)^2}{(\varepsilon_{\bf q} + \omega_{\alpha, {\bf q}})^3} + \frac{1 - \cos(q_i \, a)}{(\varepsilon_{\bf q} + \omega_{\alpha, {\bf q}})^2} \right] \nonumber \\
    &\quad - \frac{1}{2 \, d} \frac{U^2_{12}}{I^2} \mathlarger{\sum_{\alpha, \beta} \sum_{\bf q, q'} \sum_{i = 1}^{d}} \left| W_{\alpha {\bf q}, \beta {\bf q'}} + W_{\beta {\bf q'}, \alpha {\bf q}} \right|^2 \left[\frac{4 \, J \, \sin(\left[ q_i + q'_i \right] a)^2}{(\omega_{\alpha, {\bf q}} + \omega_{\beta, {\bf q'}} + \varepsilon_{\bf q + q'})^3} + \frac{1 - \cos(\left[ q_i + q'_i \right] a)}{(\omega_{\alpha, {\bf q}} + \omega_{\beta, {\bf q'}})^2 + \varepsilon_{\bf q + q'}} \right] \, , \label{eq:mstarqgw}
\end{align}
where we notice that the usual UV divergence encountered in the continuum is here regularized by the second term in the bracketed numerators of Eq.~\eqref{eq:mstarqgw}, scaling with natural short-distance cutoff scale $a$. The quasiparticle residue is given by
\begin{align}
          Z({\bf k})^{-1} &= \left[ 1 - \left. \frac{\partial \text{Re} \, \Sigma{\left( \mathbf{k}, \omega \right)}}{\partial \omega} \right|_{\omega = \varepsilon_{\mathbf{k}}} \right]^{-1} \\
          &= \ 1 + \frac{U^2_{12}}{I} \mathlarger{\sum_\alpha \sum_{\bf q}} \frac{|N_{\alpha, {\bf q}}|^2}{(\varepsilon_{\bf k} - \varepsilon_{\bf k + q} - \omega_{\alpha, {\bf q}})^2} + \frac{U^2_{12}}{2 \, I^2} \mathlarger{\sum_{\alpha, \beta} \sum_{\bf q, q'}} \frac{|W_{\alpha {\bf q}, \beta {\bf q'}} + W_{\beta {\bf q'}, \alpha {\bf q}}|^2}{(\varepsilon_{\bf k} - \varepsilon_{\bf k - q - q'} - \omega_{\alpha, {\bf q}} - \omega_{\alpha, {\bf q'}})^2} \ .
\end{align}
The decay rate can be obtained instead from the imaginary part of Eq.~\eqref{eq:dysonapprox}, resulting in
\begin{align}
        \Gamma(\mathbf{k}) &= -2 \, \text{Im} \, \Sigma{\left( \mathbf{k}, \varepsilon_{\mathbf{k}} \right)},\\
        &= \frac{2 \, \pi \, U^2_{12}}{I} \mathlarger{\sum_{\alpha} \sum_{\bf q}} \left| N_{\alpha, \mathbf{q}} \right|^2 \delta{\left(\varepsilon_{\mathbf{k}} - \varepsilon_{\mathbf{k + q}} - \omega_{\alpha, \mathbf{q}} \right)} \nonumber \\
        &\quad + \frac{\pi \, U^2_{12}}{I^2} \mathlarger{\sum_{\alpha, \beta}\sum_{\bf q,q'}} \left| W_{\alpha{\bf q},\beta{\bf q'}} + W_{\beta{\bf q'},\alpha{\bf q}} \right|^2\delta{\left( \varepsilon_{\mathbf{k}} - \varepsilon_{\mathbf{k + q + q'}} - \omega_{\alpha, \mathbf{q}} - \omega_{\beta, {\bf q'}} \right)} \ , \label{eq:decayrate}
\end{align}
whose resemblance with the Fermi golden rule, noted in the main text, is made now explicitly clear with $\delta$-functions describing the one- and two-particle emission processes.

\subsection{Comparing predictions of Fr\"ohlich and beyond-Fr\"ohlich models}\label{sm_subsec:comparison}
\begin{figure}[!ht]
\centering
\includegraphics[scale=0.48]{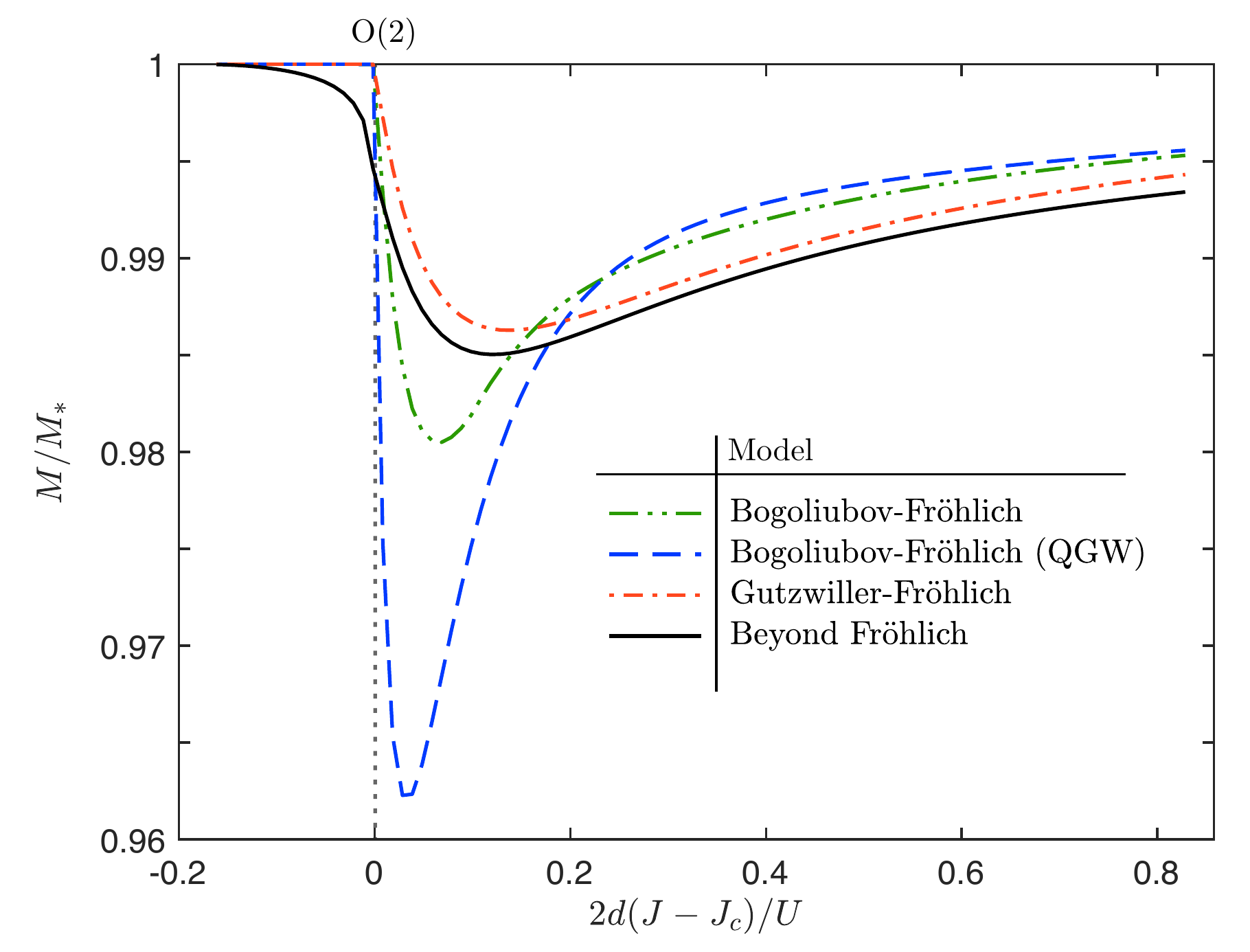}
\caption{Comparison between the results for the effective mass of the polaron across the $O(2)$ transition within different models for fixed $\langle \hat{n} \rangle = 1$. The location of the corresponding phase transitions are indicated by vertical dotted lines.}
\label{fig:SM1}
\end{figure}
In the main text, results are shown only for the beyond-Fr\"ohlich model, however it is important to understand where the Fr\"ohlich physical scenario suffices and the usual Bogoliubov treatment discussed in Sec.~\ref{sm_subsec:frohlich_models} can be used. In Fig.~\ref{fig:SM1}, we show a comparison between the predictions of different models of the bath-impurity interaction for the polaron effective mass across the $O(2)$ critical point. We immediately observe that all the approaches agree in the deep SF limit, where the effective mass reaches its bare value. Near strongly correlated regimes of the bath, the Bogoliubov-Fr\"olich result [dot-dashed green line] begins to deviate from its Gutzwiller reformulation\footnote{See the related discussion in Sec.~\ref{sm_subsec:frohlich_models}.} [dashed blue line], with the latter giving a heavier effective mass as a consequence of including the contribution of a larger number of excitation branches and stronger interactions. However, we point out that in general the Bogoliubov-Fr\"olich theory tends to overestimate significantly the weight of quantum fluctuations in the strongly-interacting SF phase, even if all the excitations on top of the condensate are considered. Instead, in this regime both Gutzwiller-Fr\"olich model [dot-dashed red line] and its beyond-Fr\"ohlich generalization [solid black line] predict a milder renormalization of the effective mass. In particular, the difference between these two curves reflects the increased contribution of two-body processes in the quantum critical regime involving the Goldstone-Higgs vertex on the SF side and particle-hole excitations on the MI side. Our beyond-Fr\"ohlich approach is able to account for these processes, which yield a smooth evolution of $M/M_*$ across the $O(2)$ transition. On the contrary, it is clear from Fig.~\ref{fig:SM1} that all Fr\"ohlich models predict instead a non-analytical behavior at this point, with the effective mass dropping trivially to its bare value as a consequence of the vanishing spectral weight of one-body excitations of the collective modes. While this non-analyticity is smeared by two-particle processes across the $O(2)$ transition, the decreased contribution of these processes across the line of CI transitions leads to increasingly narrow, non-analytic behaviors of the polaron properties [see Fig.~(3)(a) and (b)]. These findings demonstrate that the QGW treatment of the microscopic bath-impurity coupling is essential to the overall physical consistency and accuracy of the predictions presented in this work. Ultimately, the Bogoliubov scheme is reliable only in the deep superfluid regime, where the negligible depletion of the condensate justifies the corresponding expansion.

\subsection{Additional beyond-Fr\"ohlich results for the quasiparticle
residue at fixed chemical potential}\label{sm_subsec:Z_fixed_mu}
\begin{figure}[!ht]
\centering
\includegraphics[scale=0.48]{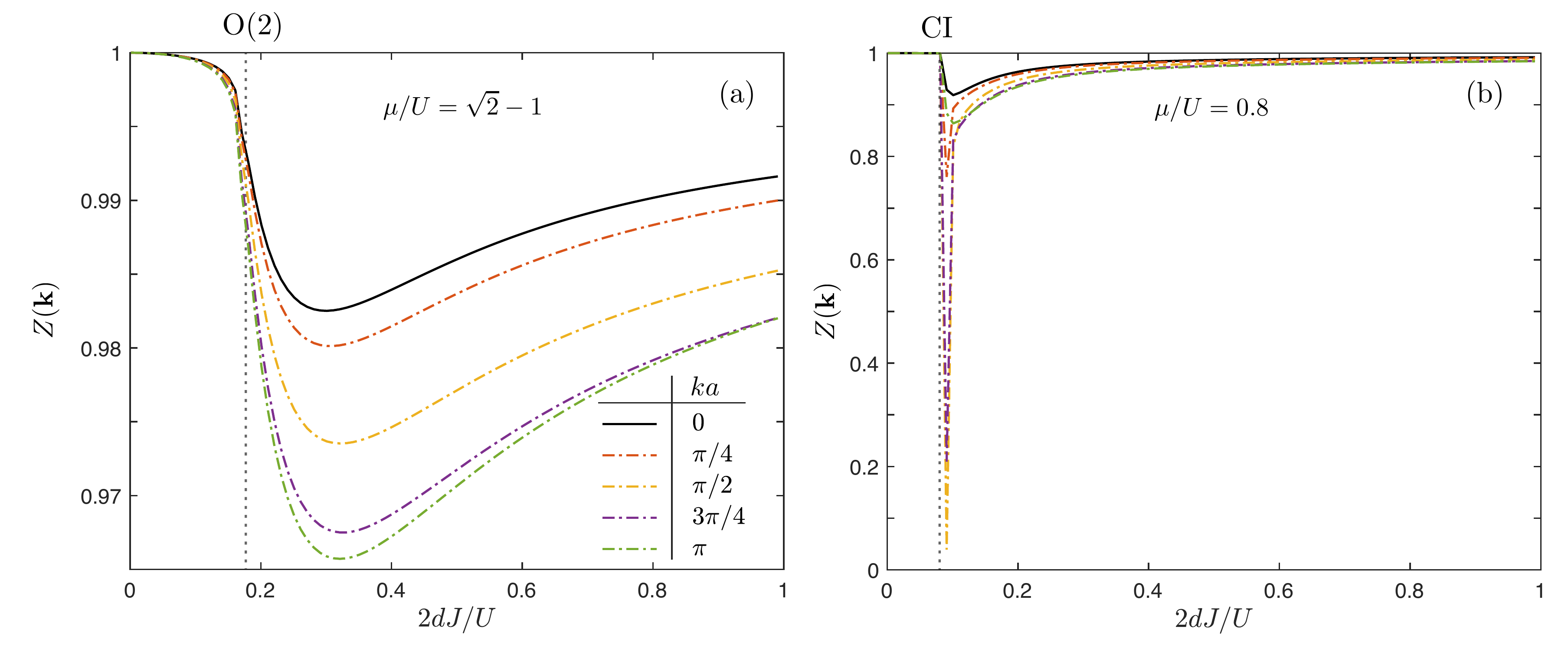}
\caption{Momentum dependence of the quasiparticle residue $Z({\bf k})$ for fixed (a) $\mu/U=\sqrt{2}-1$ and (b) $\mu/U=0.8$ across the $O(2)$ and CI transitions, respectively. The location of the corresponding phase transitions are indicated by vertical dotted lines.}
\label{fig:SM2}
\end{figure}
Here we provide additional results for the behavior of the quasiparticle residue at fixed chemical potential, shown in Fig.~\ref{fig:SM2} to complement the fixed-filling data shown in Fig.~(4) of the main text. The behavior of the residue while crossing the $O(2)$ transition for fixed $\mu/U = \sqrt{2}-1$ shown in Fig.~\ref{fig:SM2}(a) can be understood analogously to the crossing at fixed $\langle \hat{n} \rangle = 1$ in Fig.~(4)(b) of the main text. On the other hand, the behavior of the residue upon crossing the CI transition for fixed $\mu/U = 0.8$ shown in Fig.~\ref{fig:SM2}(b) behaves differently from the non-integer fixed filling results in Fig.~(4)(a) and (c) of the main text. In particular, we observe that $Z$ behaves sharply in the vicinity of the CI critical point, displaying a non-analytic behavior akin to what was found in Fig.~(3)(a) and (b) of the main text for the polaron energy and effective mass. Additionally, we see that, whereas the polaron is relatively well-defined for low momenta, it rapidly becomes incoherent for finite momenta, where especially the $p = \pi/2$ region of the Brillouin Zone exhibits the orthogonality catastrophe. We also note that, for CI transitions nearer to the HC point, $Z({\bf p})$ undergoes the orthogonality catastrophe for increasingly large regions of the Brillouin zone.


\newpage

\bibliographystyle{apsrev4-1}
\bibliography{./biblio}